\documentclass{icrc2009}

\usepackage{graphicx}   
\usepackage[caption=false]{caption}    
\usepackage[font=footnotesize]{subfig} 
\usepackage{fixltx2e}
\usepackage{url}

\newcommand{\shorttitle}[1]%
{\markboth{Proceedings of the 31\MakeLowercase{$^{st}$} ICRC, {\L}\'{o}d\'{z} 2009}{#1} }
\newcommand{\etal}{\MakeLowercase{\textit{et al. }}} 

\newcommand{\pubjournal}[6] {#1, #2 {\bf #3}, #4 (#5).}

\begin{document}
\title{Large-scale Galactic diffuse gamma rays observed  with the Fermi Gamma-Ray Space Telescope  }

\author{\IEEEauthorblockN{A.W. Strong\IEEEauthorrefmark{1}, on behalf of the Fermi-LAT Collaboration}

                            \\
\IEEEauthorblockA{\IEEEauthorrefmark{1}Max-Planck-Institut f\"ur extraterrestrische Physik, Garching, Germany}}

\shorttitle{Strong \etal Diffuse gamma rays with Fermi}
\maketitle

\begin{abstract}
The LAT instrument on the  Fermi Gamma-Ray Space Telescope is performing an all-sky  survey from 20 MeV to 300 GeV with unprecedented statistics and angular resolution.
This is providing a wealth of new information on the non-thermal emission from the Galactic interstellar medium with  implications for cosmic rays and Galactic structure.
First results at intermediate latitudes have already shown  good agreement with predictions based on direct measurements of cosmic rays, suggesting that at least the local (within about 1 kpc from the Sun) gamma-ray emission is understood.
We will  present the first  spectra from regions over the sky using the LAT data, and profiles for selected energies.
The aim here is to evaluate the agreement with  the models and assess what we can expect to learn as this analysis matures.\\

\end{abstract}



\section{Introduction}
  The LAT (Large Area Telescope) on the  Fermi Gamma-ray Space Telescope, launched in June 2008,  is performing an all-sky gamma-ray survey from 20 MeV to $>$300 GeV with unprecedented statistics and angular resolution. The sky survey started on 4 August 2008.
It provides a wealth of new information on the non-thermal emission from the Galactic interstellar medium with its implications for cosmic rays and Galactic structure.
First results at intermediate latitudes have already shown a good agreement with predictions based on direct measurements of cosmic rays \cite{Abdo09a,PorterICRCa}, suggesting that at least the local emission is understood in terms of known interstellar emission processes.
At the conference  we will present the first evaluation of the full gamma-ray sky,
comparing with appropriate models for the emission.
The aim  is to evaluate the agreement with or deviations from the models and assess what we can expect to learn as this analysis matures.

\section{Data}
We will  use data from the first year  of the mission. For some information on the data processing see \cite{Abdo09a,PorterICRCa}, and the conference presentations. Details of the LAT instrument can be found in \cite{Atwood09}, and the on-orbit calibration is described in \cite{Abdo09b}.
\section{Modelling  the Gamma-Ray Sky}
The Fermi-LAT collaboration has adopted the GALPROP cosmic-ray propagation code \cite{SM98,SMR04,SMRRD04,Ptuskin06,AnnRev57,Strong2009} for modelling the Galactic interstellar gamma-ray emission;
models of diffuse gamma-ray emission calculated with GALPROP are used also for Fermi-LAT source analysis, for example for the Bright Source List \cite{Abdo09c}.
The inputs to GALPROP used here start from the so-called `conventional model', using locally-measured cosmic-ray fluxes, and described in  \cite{SMR04}.
The components of the diffuse emission model are: pion-decay from interactions of cosmic-ray (CR) hadrons with interstellar gas, bremsstrahlung from interactions of CR electrons and positrons with interstellar gas, inverse Compton emission from CR electrons and positrons scattering off the interstellar radiation field, isotropic background, and detected point sources. 
The electron spectrum takes into account  the the Fermi-LAT measurements \cite{Abdo09d}.
The calculation includes secondary electrons and positrons which make a non-negligible contribution to inverse Compton and bremsstrahlung emission.
Full details of the model will be given elsewhere.


\section{Topics to be addressed}

We will show spectra  for various sky regions and profiles for selected energies, comparing with models both {\it a priori} and tuned to Fermi-LAT data.
 The data shown will extend from  100 MeV to over 100 GeV.
We will illustrate the validity of the model with  fits to the data, including a scaling of the emission from atomic and molecular gas  (via pion-decay + bremsstrahlung) and inverse Compton emission,
 as well as the isotropic component (instrumental + extragalactic).
The main conclusions obtained so far will be highlighted. 
These will include: the Galactic distribution and spectrum of cosmic-ray sources, the molecular hydrogen content of the Galaxy and its relation to its tracer CO molecule,
the contribution of emission processes in the interstellar medium, and the extent of the Galactic gamma-ray halo.
\section{Other Fermi-LAT analyses of diffuse emission}
In addition to the comparisons with the  models described above, a number of complementary analyses are in progress within the Fermi-LAT collaboration.
 These include a local HI correlation analysis, emission at intermediate Galactic latitudes \cite{PorterICRCa}, a study of the Orion molecular clouds \cite{OkumuraICRC}, a spatial analysis of the 2nd Galactic quadrant \cite{TibaldoICRC}, emission from the Large Magellanic Cloud \cite{PorterICRCb}
and work towards deriving the  extragalactic diffuse background \cite{AckermannICRC}. A number of these topics will be presented at this conference.

\section{Acknowledgments}

The $Fermi$ LAT Collaboration acknowledges support from a number of agencies and institutes for both development and the operation of the LAT as well as scientific data analysis. These include NASA and DOE in the United States, CEA/Irfu and IN2P3/CNRS in France, ASI and INFN in Italy, MEXT, KEK, and JAXA in Japan, and the K.~A.~Wallenberg Foundation, the Swedish Research Council and the National Space Board in Sweden. Additional support from INAF in Italy for science analysis during the operations phase is also gratefully acknowledged.

\newpage 

\end{document}